\documentclass[12pt]{article}
\usepackage{epsf}
\def\hybrid{\topmargin 0pt      \oddsidemargin 0pt
	\headheight 0pt \headsep 0pt
	\textheight 9in         
	\textwidth 6.25in       
	\marginparwidth .875in
	\parskip 5pt plus 1pt   \jot = 1.5ex}

\catcode`\@=11
\def\marginnote#1{}
\newcount\hour
\newcount\minute
\newtoks\amorpm
\hour=\time\divide\hour by60
\minute=\time{\multiply\hour by60 \global\advance\minute by-\hour}
\edef\standardtime{{\ifnum\hour<12 \global\amorpm={am}%
	\else\global\amorpm={pm}\advance\hour by-12 \fi
	\ifnum\hour=0 \hour=12 \fi
	\number\hour:\ifnum\minute<10 0\fi\number\minute\the\amorpm}}
\edef\militarytime{\number\hour:\ifnum\minute<10 0\fi\number\minute}

\def\draftlabel#1{{\@bsphack\if@filesw {\let\thepage\relax
   \xdef\@gtempa{\write\@auxout{\string
      \newlabel{#1}{{\@currentlabel}{\thepage}}}}}\@gtempa
   \if@nobreak \ifvmode\nobreak\fi\fi\fi\@esphack}
	\gdef\@eqnlabel{#1}}
\def\@eqnlabel{}
\def\@vacuum{}
\def\draftmarginnote#1{\marginpar{\raggedright\scriptsize\tt#1}}

\def\draft{\oddsidemargin -.5truein
	\def\@oddfoot{\sl preliminary draft \hfil
	\rm\thepage\hfil\sl\today\quad\militarytime}
	\let\@evenfoot\@oddfoot \overfullrule 3pt
	\let\label=\draftlabel
\let\marginnote=\draftmarginnote
	\let\marginnote=\draftmarginnote
   \def\@eqnnum{(\theequation)\rlap{\kern\marginparsep\tt\@eqnlabel}%
\global\let\@eqnlabel\@vacuum}  }

\def\numberbysection{\@addtoreset{equation}{section}
	\def\theequation{\thesection.\arabic{equation}}}

\def\underline#1{\relax\ifmmode\@@underline#1\else
	$\@@underline{\hbox{#1}}$\relax\fi}

\def\titlepage{\@restonecolfalse\if@twocolumn\@restonecoltrue\onecolumn
     \else \newpage \fi \thispagestyle{empty}\c@page\z@
	\def\thefootnote{\fnsymbol{footnote}} }

\def\endtitlepage{\if@restonecol\twocolumn \else  \fi
	\def\thefootnote{\arabic{footnote}}
	\setcounter{footnote}{0}}  
\catcode`@=12
\relax

\def\beq{\begin{equation}}
\def\eeq{\end{equation}}
\def\bea{\begin{eqnarray}}
\def\eea{\end{eqnarray}}
\def\nn{\nonumber}

\relax
\hybrid

\begin{document}

\begin{titlepage}
\begin{center}
January~2005 \hfill . \\[.5in]
{\large\bf The third parafermionic chiral algebra with the symmetry $Z_{3}$} 
\\[.5in] 
{\bf Vladimir S.~Dotsenko\bf${}^{(1)}$ 
 \bf and Raoul Santachiara\bf${}^{(2)}$}\\[.2in]
{\bf (1)} {\it LPTHE\/}\footnote{Laboratoire associ\'e No. 280 au CNRS},
         {\it Universit{\'e} Pierre et Marie Curie, Paris VI\\
               Bo\^{\i}te 126, Tour 16, 1$^{\it er}$ {\'e}tage,
               4 place Jussieu, F-75252 Paris Cedex 05, France.}\\
\hspace{-0.5cm}{\bf (2)} {\it  ITFA, Instituut voor Theoretische Fysica \\
               Valckenierstraat 65, 1018 XE Amsterdam, The Netherlands},

\end{center}

\underline{Abstract.}

We have constructed the parafermionic chiral algebra with the principal
parafermionic fields $\Psi,\Psi^{+}$ having the conformal dimension
$\Delta_{\Psi}=8/3$ and realizing the symmetry $Z_{3}$.

\end{titlepage}

\newpage

Most widely known is the first series of $Z_{N}$ parafermionic theories, which
has been defined in [1]. Its numerous applications and appearances are well known.
Less known is the second $Z_{N}$ series. Its chiral algebra has also been found
in [1], while its representation theory has been defined rather recently, in [2].
For the particular case of $N=3$, i.e. for the second $Z_{3}$ parafermionic algebra,
the corresponding conformal field theory has been developed long ago, in [3].

Restricting our discussion to the $Z_{3}$ case, the first and the
second $Z_{3}$ parafermions have respectively 
conformal dimension $\Delta_{\Psi}=2/3$ and $\Delta_{\Psi}=4/3$.
More generally, the associativity constraints (a small fraction of them)  allow for
the following discrete set of values of $\Delta_{\Psi}$, for $Z_{3}$ parafermions:
\beq
\Delta_{\Psi}=\frac{2}{3},\frac{4}{3},\frac{8}{3},\frac{10}{3},\mbox{etc.}=
\frac{2l}{3},\,\,\,l=1,2,4,5,\mbox{etc.}
\eeq
Classified in this order, in this paper we shall present the third solution of the
associativity constraints: its chiral algebra, with the basic parafermions having
the conformal dimension $8/3$.

In this short communication we intend to present our main results on the third
$Z_{3}$ parafermions. More details and demonstrations will be given in [4].

In the cases of the first and the second parafermions, the chiral 
 algebra is made of 
 $\Psi,\Psi^{+}$ and $T(z)$, the Virasoro part. The first main
 difference of the  
third $Z_{3}$ parafermions is that, in addition to the principal couple of chiral fields
$\Psi(z),\Psi^{+}(z)$ with $\Delta_{\Psi}=\Delta_{\Psi^{+}}=8/3$, the chiral algebra
of local fields contains also the fields $\tilde{\Psi}(z),\tilde{\Psi}^{+}(z)$
with the dimension $\Delta_{\tilde{\Psi}}=\Delta_{\tilde{\Psi}^{+}}=\Delta_{\Psi}+2$,
and one boson field $B(z)$ with $\Delta_{B}=4$, plus as usual $T(z)$.

This algebra is of the following form:
\bea
\Psi(z')\Psi(z)=\frac{1}{(z'-z)^{\Delta_{\Psi}}}\{\lambda\Psi^{+}(z)+(z'-z)\lambda
\beta^{(1)}_{\Psi\Psi,\Psi^{+}}\partial\Psi^{+}(z)\nn\\+(z'-z)^{2}[\lambda\beta^{(11)}
_{\Psi\Psi,\Psi^{+}}\partial^{2}\Psi^{+}(z)+\lambda\beta^{(2)}_{\Psi\Psi,\Psi^{+}}
L_{-2}\Psi^{+}(z)+\zeta\tilde{\Psi}(z)]+...\}
\eea
\bea
\Psi(z')\Psi^{+}(z)=\frac{1}{(z'-z)^{2\Delta_{\Psi}}}\{1+(z'-z)^{2}\frac{2\Delta}
{c}T(z)+(z'-z)^{3}\frac{\Delta}{c}\partial T(z)\nn\\
+(z'-z)^{4}[\beta^{(112)}_{\Psi\Psi^{+},I}\partial^{2}T(z)+\beta^{(22)}_{\Psi\Psi^{+},
I}\Lambda(z)+\gamma B(z)]\nn\\
+(z'-z)^{5}[\beta^{(1112)}\partial^{3}T(z)+\beta^{(122)}\partial\Lambda(z)+\gamma
\beta^{(1)}_{\Psi\Psi^{+},B}\partial B(z)]+...\}
\eea
\bea
\Psi(z')\tilde{\Psi}(z)=\frac{1}{(z'-z)^{\Delta_{\Psi}+2}}\{\zeta\Psi^{+}(z)+
(z'-z)^{2}[\zeta\beta^{(11)}_{\Psi\tilde{\Psi},\Psi^{+}}\partial^{2}\Psi^{+}(z)\nn\\+
\zeta\beta^{(2)}_{\Psi\tilde{\Psi},\Psi^{+}}L_{-2}\Psi(z)+
\eta\tilde{\Psi}^{+}(z)]
+(z'-z)^{3}[\zeta\beta^{(111)}_{\Psi\tilde{\Psi},\Psi^{+}}\partial^{3}\Psi(z)+
\zeta\beta^{(12)}_{\Psi\tilde{\Psi},\Psi^{+}}\partial L_{-2}\Psi(z)\nn\\+\zeta
\beta^{(3)}_{\Psi\tilde{\Psi},\Psi^{+}}L_{-3}\Psi(z)+\eta\beta^{(1)}_{\Psi\tilde
{\Psi},\tilde{\Psi}^{+}}\partial\tilde{\Psi}^{+}(z)]+...\}
\eea
\beq
\Psi(z')\tilde{\Psi}^{+}(z)=\frac{1}{(z'-z)^{2\Delta_{\Psi}-2}}\{\mu B(z)+(z'-z)\mu
\beta^{(1)}_{\Psi\tilde{\Psi}^{+},B}\partial B(z)+...\}
\eeq
\bea
\Psi(z')B(z)=\frac{1}{(z'-z)^{4}}\{\gamma\Psi(z)+(z'-z)\gamma\beta^{(1)}_{\Psi B,
\Psi}\partial\Psi(z)+(z'-z)^{2}[\gamma\beta^{(11)}_{\Psi B,\Psi}\partial^{2}
\Psi(z)\nn\\+\gamma\beta^{(2)}_{\Psi B,\Psi}L_{-2}\Psi(z)+\mu\tilde{\Psi}(z)]
+(z'-z)^{3}[\gamma\beta^{(111)}_{\Psi B,\Psi}\partial^{3}\Psi(z)\nn\\
+\gamma\beta^{(12)}
_{\Psi B,\Psi}\partial L_{-2}\Psi(z)+\gamma\beta^{(3)}_{\Psi B,\Psi} L_{-3}
\Psi(z)+\mu\beta^{(1)}_{\Psi B,\Psi}\partial\tilde{\Psi}(z)]+...\},
\eea
where the $L_{n}$ represent the modes of $T(z)$ and the coefficients
$\beta_{..}^{..}$  are fixed by the conformal symmetry.

Besides  this principal products, which are actually
sufficient to address the calculus of representations, one must consider  
also the following expansions:
\begin{itemize}
\item $\tilde{\Psi}\times\tilde{\Psi}\sim \eta
  [\Psi]+\tilde{\lambda}[\tilde{\Psi}$]. Here and in the following,
  the notation $[A]$ indicates the operator $A$ and its Virasoro descendants. 
\item $\tilde{\Psi}\times\tilde{\Psi}^{+}\sim [I]
  +\tilde{\gamma}[B(z)]$, with $I$ the identity operator. 
\item
$\tilde{\Psi}\times B\sim \mu[\Psi]+ \tilde
{\gamma}[\tilde{\Psi}]$. 
\item $B\times B\sim [I] +\xi [B]$. 
\end{itemize}
The Virasoro content of these developments is standard, containing the
(Virasoro) descendants of the above mentioned chiral fields with the corresponding
$\beta$ coefficients. Two rules have to be respected everywhere: these
developments should go (i.e. made explicit in terms of the above mentioned fields)
only up to the operators of the fifth level in the $Z_{3}$ neutral sector (which
corresponds to the number of singular terms in (3)), and up to the operators of
the third level in the $Z_{3}$ charged sector. In (2), the third level operators are 
not shown because in the case of the symmetric product $(\Psi\Psi)$ the third level
terms are uniquely defined in terms of the derivatives of the preceding levels operators;
 they do not provide new relations in the eventual calculations of the highest
weight representations. They could be added in (2), but in this
case they are irrelevant.

One could verify that, with the above prescriptions concerning  
the explicit terms  in  the 
equations (2)-(6), the corresponding representation theory is well
defined: one could define all the matrix elements needed in the degeneracy
calculations [4].

There are a total of 8 coupling constants entering into the equations
(2)-(6) and into the 
four expansions listed above. They could be
defined  through the following three-point functions:
\bea
<\Psi\Psi\Psi>=\lambda,\quad <\Psi\Psi\tilde{\Psi}>=\zeta,\quad <\Psi\Psi^{+}
B>=\gamma\nn\\
<\Psi\tilde{\Psi}\tilde{\Psi}>=\eta,\quad <\tilde{\Psi}\tilde{\Psi}\tilde{\Psi}>
=\tilde{\lambda},\quad <\Psi\tilde{\Psi}^{+}B>=\mu\nn\\
<\tilde{\Psi}\tilde{\Psi}^{+}B>=\tilde{\gamma},\quad <BBB>=\xi
\eea
Here $<\Psi\Psi\Psi>=<\Psi(\infty)\Psi(1)\Psi(0)>$ etc. .

The associativity constraints fix these constants 
in terms of the Virasoro algebra central charge $c$ which remains a free 
parameter. The results are given below:
\beq
\lambda=\frac{14}{27}\sqrt{3}\sqrt{\frac{c+32}{c}}
\eeq
\beq
\zeta=\frac{8}{27}\sqrt{30}\sqrt{\frac{(c+56)(11c+14)}{c(784+57c)}}
\eeq
\beq
\gamma=\frac{4}{27}\sqrt{15}\sqrt{\frac{(c+56)(11c+14)}{c(22+5c)}}
\eeq
\beq
\eta=\frac{7}{27}\frac{(349c+2688)\sqrt{3}}{784+57c}\sqrt{\frac{c+32}{c}}
\eeq
\beq
\tilde{\lambda}=\frac{98}{135}\sqrt{30}\frac{(c+32)(4877c^{2}+51466c+13104)}
{\sqrt{c(c+56)(11c+14)(784+57c)^3}}
\eeq
\beq
\mu=\frac{28}{9}\sqrt{6}\sqrt{\frac{(c+32)(22+5c)}{c(784+57c)}}
\eeq
\beq
\tilde{\gamma}=\frac{7}{135}\sqrt{15}\frac{(20595c^{3}+823534c^2+5532912c+2121728)}
{\sqrt{c(22+5c)(c+56)(11c+14)}(784+57c)}
\eeq
\beq
\xi=\frac{2}{15}\sqrt{15}\frac{(85c^{2}+2566c+976)}{\sqrt{c(22+5c)(c+56)(11c+14)}}
\eeq
Some details on the techniques used to obtain theses results will 
be given in \cite{ref4}.

It could be observed that for $c=-14/11$ this $Z_{3}$ algebra closes by just
$\Psi,\Psi^{+}$ and $T$, while the fields
$\tilde{\Psi},\tilde{\Psi}^{+},B$ decouple from the theory.
At this particular value of $c$, the fields $\Psi,\Psi^{+}$ belong to the Virasoro
non-unitary minimal model with $p'/p=11/6$, $\Psi,\Psi^{+}\sim\Phi_{1,3}$.

Certain analogies  with the $Z_{2}$ symmetric chiral algebras could also be mentioned
here. The allowed dimensions for the $\Psi$ operators, which realize the
$Z_{2}$ multiplication rules (and which are fermionic in this case) are the following:
\beq
\Delta_{\Psi}=\frac{1}{2},\frac{3}{2},\frac{5}{2},\frac{7}{2}...
\eeq
 Again, classified in this way by the sequence
of the dimensions $\Delta_{\Psi}$, the first solution for the associative chiral 
algebra with $Z_{2}$ symmetry corresponds to free fermions
(and the Ising model). The second solution, with $\Delta_{\Psi}=\frac{3}{2}$,
corresponds to the $N=1$ superconformal algebra, which, in a sense, is just a less
trivial realization of the $Z_{2}$ symmetry, as compared to free fermions.

The third solution, with $\Delta_{\Psi}=5/2$, is know as the chiral algebra of the
$WB_{2}$ model [5]. Starting with the third solution, the associativity of the
chiral algebra of the $\Psi$ field and its closure (while preserving $c$ as a free
parameter) requires the introduction of extra fields: $B(z)$ chiral field with
$\Delta_{B}=4$ in case of $\Delta_{\Psi}=5/2$; more bosonic fields for still
higher solutions in the sequence (16) (they are the models $WB_{n}$ [5]).

In general, when the value of $\Delta_{\Psi}$ grows, the number of singular terms
in the expansions of $\Psi(z')\Psi(z)$ and $\Psi(z')\Psi^{+}(z)$, which have to be
defined explicitly, grows also. The gap becomes too big to be filled by just
Virasoro descendants of $\Psi,\Psi^{+}$ and to satisfy associativity.

The appearance of extra conserved currents for higher solutions has
also a 
physical meaning. Indeed, for a given symmetry,  higher solutions are
likely to correspond to higher multicriticalities. 
Naturally, the number of degrees of freedom of the corresponding
(multi)critical theories should grow.

\underline{Acknowledgments:} We are grateful to J.L.Jacobsen for his collaboration
during the initial stages of this project.

\end{document}